\documentclass{ws-ijmpa}
\usepackage[super,compress]{cite}
\usepackage{graphicx}
\begin{document}
\markboth{Masanori Hanada}{What lattice theorists can do for quantum gravity}
\newcommand{\Slash}[1]{\ooalign{\hfil/\hfil\crcr$#1$}}

\catchline{}{}{}{}{}
%
%%%%%%%%%%%%%%%%%%%%%%%%%%%%%%%%%%%%%%%%%%%%%%%%%%%%%%%%%%%%%%%%%%%%

\title{What lattice theorists can do for superstring/M-theory}

\author{MASANORI HANADA%\footnote{
%Typeset names in 8 pt roman, uppercase. Use the footnote to indicate the
%present or permanent address of the author.}
}

\address{
Yukawa Institute for Theoretical Physics\\
Kyoto University, Kyoto 606-8502, JAPAN\\
The Hakubi Center for Advanced Research \\ 
Kyoto University, Kyoto 606-8501, JAPAN\\
Stanford Institute for Theoretical Physics\\
Stanford University, Stanford, CA 94305, USA\\
hanada@yukawa.kyoto-u.ac.jp}

\maketitle

%\begin{history}
%\received{Day Month Year}
%\revised{Day Month Year}
%\end{history}

\begin{abstract}

The gauge/gravity duality provides us with nonperturbative formulation of superstring/M-theory. 
Although inputs from gauge theory side are crucial for answering many deep questions 
associated with quantum gravitational aspects of superstring/M-theory, 
many of the important problems have evaded analytic approaches. 
For them, lattice gauge theory is the only hope at this moment. 
In this review I give a list of such problems, putting emphasis on problems within reach in a five-year span, 
including both Euclidean and real-time simulations.

\keywords{Lattice Gauge Theory; Gauge/Gravity Duality; Quantum Gravity.}
\end{abstract}

\ccode{PACS numbers:11.15.Ha,11.25.Tq,11.25.Yb,11.30.Pb}
\ccode{Preprint number:YITP-16-28}
%\tableofcontents

%%%%%%%%%%%%%%%%%
%%%%%%%%%%%%%%%%%
\section{Introduction}	
%%%%%%%%%%%%%%%%%
%%%%%%%%%%%%%%%%%
During the second superstring revolution, string theorists built a beautiful framework for quantum gravity: the gauge/gravity duality\cite{Maldacena:1997re,Itzhaki:1998dd}. 
It claims that superstring/M-theory is equivalent to certain gauge theories, at least about certain background spacetimes 
(e.g. black brane background, asymptotically anti de-Sitter (AdS) spacetime). 
Here the word `equivalence' would be a little bit misleading, because it is not clear how to define string/M-theory nonperturbatively; 
superstring theory has been formulated based on perturbative expansions, and M-theory\cite{Witten:1995ex} 
is defined as the strong coupling limit of type IIA superstring theory, where the nonperturbative effects should play important roles. 
Therefore, gauge theories should be interpreted as {\it definitions} of superstring/M-theory, 
just in the same way that lattice QCD defines QCD. 

Two decades after the revolution, however, the power of the duality has not been fully utilized yet. 
The reason is simple: it is difficult to solve gauge theories. 
String theorists do not have practical tools to solve them, 
unless one considers special sectors with strong kinematical constraints, such as the integrable sector 
in the planar limit\cite{Beisert:2010jr}. 
Although such highly constrained sectors played important roles in 
testing the duality (first of all, that both gauge and gravity sides have corresponding highly constrained sectors 
is already nontrivial evidence of the duality), in order to learn about the dynamics of superstring/M-theory, 
more powerful tools applicable to generic situations are needed. 
Apparently, lattice theorists can provide such tools.  
There are many important problems which require close collaborations between lattice and string theorists, for example -- 
\begin{itemize}
\item
Schwarzschild black hole from a quantum theory, which is the key to the complete solution of the black hole information puzzle. 

\item
How stringy effects resolve the curvature singularity of general theory of relativity.  

\item
The quantum nature of M-theory, or first of all -- what is M-theory?

\item
Holographic descriptions of cosmology. 

\item
Six-dimensional ${\cal N}=(2,0)$ super conformal field theory, which is mother to various field theory dualities. 

\end{itemize}

In this review, I will explain main ideas, without going into too many technical details. 
In Sec.~\ref{sec:duality}, I briefly introduce gauge/gravity duality. 
Rather than explaining generic features of the duality, I will show explicit examples, 
which can be studied by Monte Carlo methods in the near future. 
In Sec.~\ref{sec:imaginary_time_simulation} I introduce regularization methods which can be used for Euclidean simulations. 
I will also mention few details on actual simulations (Rational Hybrid Monte Carlo (RHMC) algorithm, sign problem and flat directions), 
which usual lattice QCD practitioner might not be familiar with. 
A part of previous simulation results will also be explained. 
Sec.~\ref{sec:real_time} explains the real-time simulation. 
Rather interestingly, it is much easier compared to QCD, 
in the sense that the weak coupling region already has valuable implications to quantum gravity. 

%%%%%%%%%%%%%%%%%
%%%%%%%%%%%%%%%%%
\section{Gauge/gravity duality}\label{sec:duality}	
%%%%%%%%%%%%%%%%%
%%%%%%%%%%%%%%%%%
Outside string community (and sometimes even among string theorists!) there are widespread misunderstandings 
about the gauge/gravity duality conjecture; 
typical misconceptions include --  
{\it the duality relates (ONLY) large-$N$ gauge theory in the strong coupling limit and classical gravity}, 
and {\it it is used (ONLY) to simplify the analysis of quantum field theory in special limits.}
This is very unfortunate for lattice theorists for two reasons: 
Firstly, such misunderstandings emerged because 
lattice theorists did not work in this field (or string theorists could not use lattice techniques), 
and hence nobody could solve hard problems on quantum field theory.  
Secondly, once such misunderstandings spread they prevented lattice theorists to work in this field. 
It was a typical negative feedback caused by only sociological reasons.  

The actual statement\cite{Maldacena:1997re,Itzhaki:1998dd} is as follows: 
Certain gauge theories 
should be equivalent to superstring theory, including stringy effect.  
If one takes large-$N$ and strong coupling limit in the gauge theory side, the gravity side reduces to supergravity 
(supersymmetric generalization of Einstein gravity). Finite-$N$ and finite-coupling corrections to gauge theory  
correspond to stringy corrections in the gravity side. 
There are many generalizations of this correspondence, including a correspondence to M-theory. 

Let us formulate the duality more precisely. As the simplest and best established example, 
let us consider  maximally supersymmetric Yang-Mills theory (maximal SYM) in $(p+1)$ spacetime dimensions with U$(N)$ gauge group.
Formally, they can be obtained from $(9+1)$-dimensional ${\cal N}=1$ SYM via the dimensional reduction. 
The action of the $(9+1)$-dimensional theory is (with Euclidean signature) 
\begin{eqnarray}
S_{(9+1)} = 
\int d^{10}x {\rm Tr}\left(
\frac{1}{4}F_{\mu\nu}^2 + \frac{i}{2}\bar{\psi}\Slash{D}\psi
\right), 
\end{eqnarray}
where $F_{\mu\nu}=\partial_\mu A_\nu-\partial_\nu A_\mu -i[A_\mu,A_\nu]$ is the field strength, 
$A_\mu$ ($\mu=1,2,\cdots,10$) is the gauge field which is $N\times N$ Hermitian, 
$\psi_\alpha$ ($\alpha=1,2,\cdots,16$) is an adjoint fermion in the Majorana-Weyl representation, 
on which the covariant derivative acts as $D_\mu\psi=\partial_\mu\psi-i[A_\mu,\psi]$.  
$16\times 16$ matrices $\gamma_\mu$ are the left-handed sector of ten-dimensional gamma matrices. 
This action is invariant under the supersymmetry transformation $A_\mu\to A_\mu+i\bar{\epsilon}\gamma_\mu\psi$, 
$\psi\to\psi+\frac{1}{2}F_{\mu\nu}\gamma^{\mu\nu}\epsilon$. There are 16 supercharges, corresponding to the 16 components of $\epsilon$. 
The dimensionally reduced theory in $(p+1)$ dimensions can be obtained by dropping the dependence on $x^{a}$, $a=p+2,\cdots,10$.  
 $A_a$ turns to scalar in $(p+1)$-d theory, $X_{a-p-1}$. The action becomes 
\begin{eqnarray}
S_{(p+1)} = 
\int d^{p+1}x {\rm Tr}\left(
\frac{1}{4}F_{\mu\nu}^2 
+
\frac{1}{2}(D_\mu X_I)^2
-
 \frac{1}{4}[X_I,X_J]^2 
 +
\frac{i}{2}\bar{\psi}\Slash{D}\psi
+
\frac{1}{2}\bar{\psi}\tilde{\gamma}^I [X_I,\psi]
\right), \nonumber\\
\end{eqnarray}
where $\mu,\nu=1,2,\cdots,p+1$, $I,J=1,\cdots,9-p$ and $\tilde{\gamma}^I=\gamma^{I+p+1}$. 
Obviously, it preserves 16 supercharges. 
This is `maximally supersymmetric' in the sense that, with more supercharges, fields with spin larger than 1 necessarily appear 
and hence it is impossible to write down the local, ghost-free interacting theory without including gravity. 
The connections to string theory has been realized first as an effective description\cite{Witten:1995im} 
of low-energy dynamics of D$p$-branes\cite{Dai:1989ua,Polchinski:1995mt} and open strings. 

The  D$p$-brane  is  an  object  extending  to  the  time and $p$ spatial dimensions. 
While the closed string can propagate anywhere in the bulk
$(1+9)$-dimensional spacetime, the endpoints of the open string must be attached to the D-brane,
or equivalently the Dirichlet (Neumann) boundary condition along
$x^{p+2},\cdots,x^{10}$ ($x^1,\cdots,x^{p+1}$)
is imposed for the endpoints. 
The spacetime of the gauge theory $(x_1,x_2,\cdots,x_{p+1})$ corresponds to the D-brane world volume, 
and diagonal elements $(X_1^{ii},X_2^{ii},\cdots,X_{9-p}^{ii})$ describes the position of $i$-th D$p$-brane along the 
transverse directions ($x^{p+2},\cdots,x^{10}$).
The off-diagonal elements $X_I^{ij}$ describes open string excitations between $i$-th and $j$-th D$p$-branes. 
(For more precise interpretation, see e.g. Ref.\citen{Azeyanagi:2009zf}. For an elementary review, see Ref.\citen{Hanada:2012eg}.)
In type IIA and IIB superstring theories, D$p$-brane with even and odd values of $p$ can exist as stable, supersymmetric objects, respectively. 

D-branes are massive objects. Therefore, when many D-branes are put on top of each other, 
the geometry gets curved and becomes the {\it black brane}, which is a higher-dimensional analogue of the black hole. 
The metric of the black $p$-brane is given by \cite{Gibbons:1987ps,Horowitz:1991cd}  
\begin{eqnarray}
ds^2
&=&
\alpha'\Biggl\{
\frac{U^{(7-p)/2}}{g_{YM}\sqrt{d_p N}}
\left[
-\left(
1-\frac{U_0^{7-p}}{U^{7-p}}
\right)dt^2
+
dy_{\parallel}^2
\right]
\nonumber\\
& &
\qquad
+
\frac{g_{YM}\sqrt{d_p N}}{U^{(7-p)/2}\left(1-\frac{U_0^{7-p}}{U^{7-p}}\right)}dU^2
+
g_{YM}\sqrt{d_p N}U^{(p-3)/2}d\Omega_{8-p}^2
\Biggl\},
\label{near_extremal_metric} 
\end{eqnarray}
where we have used the Yang-Mills coupling $g_{YM}$ of the corresponding SYM for later convenience. 
A constant $\alpha'$ is the inverse of the string tension, 
$(t,y_\parallel)$ is the coordinate of the $(p+1)$-dimensions along the brane,  
and $U$ and $\Omega$ are the radial and angular coordinate of the transverse directions.  
 The horizon coordinate $U_0$ is related to the Hawking temperature of the black brane by 
\begin{eqnarray}
T=\frac{(7-p)U_0^{(5-p)/2}}{4\pi \sqrt{d_p g_{YM}^2N}},  
\end{eqnarray}
where $d_p=2^{7-2p}\pi^{(9-3p)/2}\Gamma((7-p)/2)$. 
The Hawking temperature corresponds to the temperature of the gauge theory. 
The dilaton (the string coupling constant) is given by 
\begin{eqnarray}
e^{\phi}=(2\pi)^{2-p}g_{YM}^2
\left(
\frac{d_p g_{YM}^2N}{U^{7-p}}
\right)^{\frac{3-p}{4}}. 
\label{string_coupling}
\end{eqnarray}

Maldacena conjectured that both weakly-coupled type II superstring and SYM can be good descriptions of 
$N$-coincident D$p$-branes in type II superstring theory, at least in a certain limit. 
Then $(p+1)$-dimension U$(N)$ maximal SYM
should be equivalent to type II superstring theory about black $p$-brane background, including stringy effect.  
The large-$N$ and strong coupling limit of SYM corresponds to supergravity in the gravity side. 
Finite-$N$ and finite-coupling corrections to SYM 
correspond to stringy corrections in gravity side. 
\begin{eqnarray}
\begin{array}{ccc}
N=\infty, \lambda=\infty\qquad & \longleftrightarrow &\qquad  {\rm classical\ supergravity}\\
1/\lambda\ {\rm correction}\qquad  & \longleftrightarrow &\qquad  \alpha'\ {\rm correction\ (deviation\ from\ point\ particle)}\\
1/N\ {\rm correction}\qquad  & \longleftrightarrow &\qquad  g_s\ {\rm correction\ (virtual\ loops\ of\ strings)}
\end{array}
\nonumber
\end{eqnarray}

The most important aspect of this conjecture is that, while the gauge theory side is nonperturbatively defined 
(for example, by using lattice), 
we do not know how to define the string theory side without relying on perturbative expansion.  
Therefore, this `duality' should be understood as {\it definition} of superstring theory based on gauge theory. 
Then the important roles of lattice gauge theory are obvious: 
Firstly, without using lattice, it is extremely hard to test the duality conjecture quantitatively. 
For example, the duality has been assumed to be correct at finite temperature, 
and used intensively for learning about QGP\cite{Kovtun:2004de}. 
However, if one actually tries to test the duality, one has to solve both gauge theory and gravity sides.  
Then the only tractable method is the lattice Monte Carlo simulation. 
Indeed, quantitative tests at finite temperature have been performed based on Monte Carlo studies of 
$(0+1)$-d and $(1+1)$-d theories.\cite{Anagnostopoulos:2007fw,Catterall:2008yz,Catterall:2009xn,Kadoh:2015mka,Filev:2015hia,Hanada:2008gy,Catterall:2010fx} 
Although things become even harder at stringy level,
Monte Carlo methods still work nicely.\cite{Hanada:2008ez,Hanada:2013rga} 
Secondly, in order to study quantum gravitational aspects of superstring theory, usually one has to study gauge theory side, 
because it provides us with the nonperturbative definition of the theory. Analytic tools requires strong constraints, 
which typically works in some integrable and/or supersymmetric sectors. In order to learn about the dynamics, 
one has to study more generic situations, and then numerical simulation is practically the only tool. 

In the remaining part of this section, 
we explain physics in each dimensions, putting emphasis on doable things on lattice. 
%%%%%%%%%%%%%%%%%
%%%%%%%%%%%%%%%%%
\subsection{$(3+1)$-d SYM and precision test of AdS$_5$/CFT$_4$ correspondence}	
%%%%%%%%%%%%%%%%%
%%%%%%%%%%%%%%%%%

$(3+1)$-d ${\cal N}=4$ SYM is conjectured to be dual to type IIB superstring on AdS$_5\times$S$^5$, 
which is the near-horizon limit of black 3-brane geometry. 
This theory is ultraviolet-finite and conformal\cite{Mandelstam:1982cb,Brink:1982wv}.  
The $\alpha'$ correction is small when $\lambda=g_{YM}^2N\ll 1$ and the $g_s$ correction is small when $g_{YM}^2\ll 1$. 

This is the most intensively studied example of the gauge/gravity duality, if numerical simulations are not counted. 
However, the test has been done only for quantities under good analytic control. 
Therefore, precision tests based on lattice simulation are very important. 
Here is a partial list of interesting topics: 

\begin{itemize}
\item
The duality at finite temperature. Does it really hold? 
The most basic thing to calculate is the free energy at finite temperature. 
It has been suggested\cite{Gubser:1998nz} that the free energy density takes the form of 
$F=f(\lambda)N^2T^4$ up to the $1/N$ correction, where $f(\lambda)=\pi^2/6-\lambda/4+\cdots$ 
at weak coupling\cite{Fotopoulos:1998es} and 
$f(\lambda)=\pi^2/8 + 15\pi^2\zeta(3)/64\lambda^{3/2}+\cdots$ at strong coupling\cite{Gubser:1998nz},   
as calculated from gauge and gravity sides, respectively. 
Whether such interpolation function $f(\lambda)$ actually exists is an important open problem. 
(Note that essentially the same problem has been studied numerically in the case of D0-brane theory\cite{Anagnostopoulos:2007fw,Catterall:2008yz,Catterall:2009xn,Kadoh:2015mka,Filev:2015hia,Hanada:2008ez}.)

\item
Gubser-Klebanov-Polyakov-Witten (GKPW) prescription\cite{Witten:1998qj,Gubser:1998bc} 
to calculate the gauge theory correlation functions 
from gravity side. Does it really hold even without being protected by supersymmetry? 
How big is the $1/N$-correction? 
(The generalization of GKPW to D0-brane theory\cite{Sekino:1999av} has been tested numerically\cite{Hanada:2011fq,Hanada:2009ne}.)

\item
Ryu-Takayanagi formula for the entanglement entropy\cite{Ryu:2006bv}. 
How precise is it? How big is the $1/N$-correction? Can the $1/N$-correction capture 
quantum gravitational effects in the dual geometry? 

\item
S-duality\cite{Montonen:1977sn,Witten:1978mh,Osborn:1979tq} (Montonen-Olive duality). 
4d ${\cal N}=4$ SYM is expected to be invariant under the S-duality transformation, 
which transforms the coupling constant $g_{YM}$ and theta angle $\theta$ as 
$\tau=\frac{\theta}{2\pi}+\frac{4\pi i}{g_{YM}^2}\to -\frac{1}{\tau}$. 
Numerical analysis at strong coupling should be useful for testing this duality 
beyond SUSY-protected observables. 

\end{itemize}

All these problems have huge impacts, both in quantum gravity and `applied AdS/CFT', and as far as I notice, 
these problems can be addressed only when lattice theorists and string theorists work together.

%%%%%%%%%%%%%%%%%
%%%%%%%%%%%%%%%%%
\subsection{$(2+1)$-d  SYM, ABJM theory and M-theory}\label{sec:D2/M2}	
%%%%%%%%%%%%%%%%%
%%%%%%%%%%%%%%%%%
$(2+1)$-d maximal SYM near the 't Hooft large-$N$ limit is conjectured to be dual to type IIA superstring about black two-brane geometry. 
The coupling constant has the dimension of mass, and hence the theory flows to strong coupling at long distance. 
There, the $g_s$ correction grows; and, according to the M-theory conjecture\cite{Witten:1995ex}, 
eleventh dimension, `M-theory circle,' opens up 
at strong coupling regime of type IIA superstring theory. 
D2-branes in IIA superstring should turn to M2-branes in M-theory, whose near horizon geometry is AdS$_4\times$S$^7$. 
That the dual geometry is AdS means the field theory must be conformal. It cannot be Yang-Mills theory, because 
the gauge coupling is dimensionful in $(2+1)$ dimensions. Instead, Chern-Simons-matter theory 
has been regarded as a promising candidate. 
Other features like maximal supersymmetry and the structure of the moduli (possible distributions of M2-branes) must also match.
Such a theory has been constructed by Aharony, Bergman, Jafferis and Maldacena (ABJM)\cite{Aharony:2008ug}. 
This is a $U(N)\times U(N)$ supersymmetric Chern-Simons-matter theory, whose action in Euclidean signature is given by 
\begin{eqnarray}
\lefteqn{
{\mathcal L}_{U(N)\times U(N)}
=
k\,\mathrm{Tr}\,\Biggl[
\frac12\epsilon^{\mu\nu\rho}\left(
-A_\mu\partial_\nu A_\rho-\frac{2}{3}A_\mu A_\nu A_\rho
+\tilde{A}_\mu\partial_\nu \tilde{A}_\rho
+\frac{2}{3}\tilde{A}_\mu \tilde{A}_\nu \tilde{A}_\rho
\right)}
\nonumber\\
& &\quad
+\left(-D_\mu\bar{\Phi}^\alpha D^\mu\Phi_\alpha
+i\bar{\Psi}^\alpha\Slash{D}\Psi_\alpha\right)
-i\epsilon^{\alpha\beta\gamma\delta}\Phi_\alpha\bar{\Psi}_\beta\Phi_\gamma\bar{\Psi}_\delta
+i\epsilon_{\alpha\beta\gamma\delta}\bar{\Phi}^\alpha\Psi^\beta\bar{\Phi}^\gamma\Psi^\delta
\nonumber \\
&&\quad
+i\left(
-\bar{\Psi}_\beta\Phi_\alpha\bar{\Phi}^\alpha\Psi^\beta
+\Psi_\beta\bar{\Phi}_\alpha\Phi^\alpha\bar{\Psi}^\beta
+2\bar{\Psi}_\alpha\Phi_\beta\bar{\Phi}^\alpha\Psi^\beta
-2\Psi^\beta\bar{\Phi}^\alpha\Phi_\beta\bar{\Psi}_\alpha
\right)
\nonumber\\
&&\quad
+\frac{1}{3}\left(
\Phi_\alpha\bar{\Phi}^\beta\Phi_\beta\bar{\Phi}^\gamma\Phi_\gamma\bar{\Phi}^\alpha
+\Phi_\alpha\bar{\Phi}^\alpha\Phi_\beta\bar{\Phi}^\beta\Phi_\gamma\bar{\Phi}^\gamma
\right. \nonumber \\
&~& \quad\quad \quad \left.
+4\Phi_\beta\bar{\Phi}^\alpha\Phi_\gamma\bar{\Phi}^\beta\Phi_\alpha\bar{\Phi}^\gamma
-6\Phi_\gamma\bar{\Phi}^\gamma\Phi_\beta\bar{\Phi}^\alpha\Phi_\alpha\bar{\Phi}^\beta
\right)
\Biggl] \,,
\end{eqnarray}
where $A_\mu$ and $\tilde{A}_\mu$ are $U(N)$ gauge fields, and 
$\Phi_\alpha$ and 
$\Psi_\alpha\ (\alpha=1,2,3,4)$ are 
bosonic and fermionic complex bifundamental fields, respectively. 
The Chern-Simons level $k$ is quantized to integer values.  
The infrared fixed point of $(2+1)$-d maximal SYM is conjectured to be the ABJM theory with $k=1$.
This theory has ${\cal N}=8$ supersymmetry for $k=1,2$
and ${\cal N}=6$ supersymmetry for $k\ge 3$.
It has been conjectured to be dual to M-theory 
on $AdS_4\times S^7/{\mathbb Z}_k$ for $k\ll N^{1/5}$,
and to type IIA superstring on $AdS_4\times {\mathbb C}P^3$ 
in the planar large-$N$ limit with 
the 't Hooft coupling constant $\lambda=N/k$ kept fixed. 

Regarding ABJM theory, remarkable analytic results have been obtained when the supersymmetry helps, e.g. 
Refs.\citen{Kapustin:2009kz,Drukker:2010nc,Fuji:2011km}. 
Unfortunately, a lattice formulation of the ABJM theory is not known at this moment.  
Still, at large-$N$, the regularization based on the Eguchi-Kawai equivalence\cite{Eguchi:1982nm} can work; 
see Sec.~\ref{sec:Eguchi_Kawai} for details.  
Therefore precision tests of AdS$_4$/CFT$_3$ correspondence, both in type IIA and M-theory regions, can be done. 

Another interesting feature of $(2+1)$-d maximal SYM, which turns out to be important for Monte Carlo study, 
is the Myers effect\cite{Myers:1999ps} -- with suitable background flux, D$0$-branes spread out and form 
a kind of `lattice,' which can be regarded as D$2$-brane. 
In terms of gauge theory, two-dimensional `lattice' can be embedded in scalar fields in $(0+1)$-dimensional gauge theory, 
so that $(2+1)$-dimensional theory is obtained. 
A system of $N$ D0-branes is described by U($N$) maximally supersymmetric matrix quantum mechanics (see Sec.~\ref{sec:MQM}). 
To this model, it is possible to introduce background flux without breaking supersymmetry. The action changes to the so-called 
Berenstein-Maldacena-Nastase (BMN) matrix model\cite{Berenstein:2002jq}, whose action is given by 
\eqref{eq:BMN_1}, \eqref{eq:BMN_2}.  
BMN matrix model has a fuzzy sphere solution,
which keeps full 16 supersymmetry. 
As we will see in Sec.~\ref{sec:NCYM}, fuzzy spheres generate two-dimensional noncommutative space.  
By expanding the theory around $k$-coincident 
fuzzy sphere background, $(1+2)$-dimensional super Yang-Mills is obtained. 
Interesting problems to which numerical simulations are crucial include the following:
\begin{itemize}
\item
We can take a limit in which $(1+2)$-d SYM on $S^2$ with finite radius is realized.  
According to the duality conjecture, as the temperature becomes lower, the system should go through a phase transition to 
the confinement phase, in the same way as in $(1+3)$-d SYM on $S^3$ \cite{Witten:1998zw}. 
Can we test it?

\item
Let us consider $(1+2)$-dimensional super Yang-Mills on flat space. 
As we go to even lower temperature, $g_s$ correction becomes larger, the M-theory circle opens up, 
and D2-branes turn to M2-branes. 
When the M-theory circle is small, M2-branes cannot localize along the M-theory circle. As the M-theory circle becomes larger, 
at some point M2-branes localize. This transition from D2 to M2 is known as the Gregory-Laflamme transition\cite{Gregory:1993vy}. 
Can this transition be seen? 

\item
After the Gregory-Laflamme transition, AdS$_4$/CFT$_3$ description becomes better. 
Can we understand the properties of M-theory there, especially from the $1/N$-correction? 

\end{itemize}

%%%%%%%%%%%%%%%%%
%%%%%%%%%%%%%%%%%
\subsection{$(1+1)$-d  SYM and black hole/black string phase transition}	
%%%%%%%%%%%%%%%%%
%%%%%%%%%%%%%%%%%
$(1+1)$-d maximal SYM in the strong coupling region ($g_{YM}^2\to\infty$) has been proposed as 
a description of type IIA superstring theory\cite{Dijkgraaf:1997vv}. 
(This so called `Matrix String Theory' is regarded as the compactification of 
Matrix Model of M-theory\cite{Banks:1996vh,de Wit:1988ig} and hence describes IIA rather than IIB.)
Later, the 't Hooft limit has been identified with the description of black one-brane in type IIB superstring theory\cite{Itzhaki:1998dd}, 
about the trivial vacuum ($X_M=0$). 

This theory provides us with an ideal tool for studying {\it topology change} of black hole like objects\cite{Aharony:2004ig,Aharony:2005ew}. 
To see this, let us first introduce `black string' and `black hole' in this theory. 
(Note that `black string' discussed here does not mean black one-brane.)
The key notion is the T-duality\cite{Kikkawa:1984cp} which relates type IIA and IIB superstring theories 
compactified on a circle with circumference $R$ and $\alpha'/R$, respectively. Kaluza-Klein momentum and winding number are exchanged. 

We start with $(1+1)$-d U($N$) SYM compactified on spatial circle. When the compactification radius is large, it describes 
$N$ D1-branes in type IIB superstring theory winding on spatial circle.  
When the compactification radius becomes smaller, strings winding on the circle becomes lighter and 
perturbative description in terms of type IIB superstring theory becomes worse. Then it is better to switch to type IIA description, 
by using T-duality, because the T-dual circle becomes bigger as the original circle becomes smaller. 
By T-duality, D1-branes are mapped to D0-branes\cite{Dai:1989ua}. The positions of D0-branes on the T-dual circle corresponds to 
the phase of Wilson line winding on the circle,  $W={\rm P}e^{i\oint dx A_x}$. 
If we diagonalize the Wilson line as $W={\rm diag}(e^{i\theta_1},\cdots,e^{i\theta_N})$, $\theta_i$ describes the position of 
the $i$-th D0-brane. If all $\theta_i$'s clump at the same point, they describe a {\it black hole}. 
If $\theta_i$'s spread uniformly on the unit circle, {\it uniform black string} is described. 
Other phases like non-uniform string and multi black holes would also be realized depending on the distribution of the Wilson line phases. 

Study of topology change between such objects have been motivated by the discovery of the instability of 
the black string solution in Einstein gravity\cite{Gregory:1993vy}. Discovery of other topologically nontrivial objects such as 
the black ring\cite{Emparan:2001wn} added further motivations. From the point of view of superstring theory, 
the topology change is interesting because the stringy effects are essential; when black string pinches off, 
curvature singularity shows up and Einstein gravity ceases to work. Is the $\alpha'$-correction (large-$N$, finite $\lambda$) 
sufficient to resolve the singularity? What is the role of the $g_s$ correction (finite-$N$)? SYM simulation should provide us with 
valuable insights based on the first principles. The first Monte Carlo study\cite{Catterall:2010fx} at moderate values of $N$ 
with coarse lattice gave the phase diagram consistent with expectations from gravity side.  
Eventually, in order to see the moment of the phase transition, real-time simulations will be needed. 
I will comment it in Sec.~\ref{sec:real_time}. 

%%%%%%%%%%%%%%%%%
%%%%%%%%%%%%%%%%%
\subsection{$(0+1)$-d SYM and IIA superstring/M-theory}\label{sec:MQM}	
%%%%%%%%%%%%%%%%%
%%%%%%%%%%%%%%%%%
$(0+1)$-d SYM (supersymmetric matrix quantum mechanics) has been studied numerically by several groups. 
A simulation code can be downloaded freely from my homepage\cite{MatrixModelCode}. 

Historically, this model has been introduced as a regularization of supermembrane in eleven dimensional spacetime\cite{de Wit:1988ig}. 
An interpretation as a low-energy effective theory of D0-branes and open strings has been given in Ref.\citen{Witten:1995im}. 
Then it has been interpreted as a nonperturbative formulation of M-theory in the infinite-momentum frame\cite{Banks:1996vh},  
and as a dual of type IIA superstring \cite{Itzhaki:1998dd}. 
Taking the names of people who gave the M-theory interpretation, this model is often called 
Banks-Fischler-Shenker-Susskind (BFSS) matrix model. 
Parameter regions of interest vary depending on the interpretations, as I will explain later.  
This model consists of nine $N\times N$ bosonic hermitian matrices $X_M$ ($M=1,2,\cdots,9$), sixteen 
fermionic matrices $\psi_\alpha$ ($\alpha=1,2,\cdots,16$) and the gauge field $A_t$. 
Both $X_M$ and $\psi_\alpha$ are in the adjoint representation of U$(N)$ gauge group, and the covariant derivative $D_t$ acts on them as 
$D_tX_M = \partial_t X_M -i[A_t,X_M]$ and $D_t\psi_\alpha = \partial_t\psi_\alpha -i[A_t,\psi_\alpha]$. 
The action is given by%\footnote{
%In this theory, the 't Hooft coupling $\lambda=g_{YM}^2N$ has a dimension of $({\rm mass})^3$, and can be set to 1 by proper rescaling %of time $t$ and matrices.
%In other words, all dimensionful quantities can be set dimensionless by multiplying appropriate powers of $\lambda$. 
%In this code, $\lambda=1$. 
%} 
\begin{eqnarray}
S_{BFSS}
&=&
\frac{N}{\lambda}\int dt\ {\rm Tr}\left\{
\frac{1}{2}(D_t X_M)^2 
-
\frac{1}{4}[X_M,X_N]^2 
+
i\bar{\psi}\gamma^{10}D_t\psi
-
\bar{\psi}\gamma^M[X_M,\psi] 
\right\}.
\nonumber\\
\label{eq:BFSS}   
\end{eqnarray}
%Here $\gamma^M$ ($M=1,\cdots,10$), which are $16\times 16$ upper-right block of the 10d Gamma matrices $\Gamma^{M}$.
% 
%This model is obtained by dimensionally reducing the ten-dimensional ${\cal N}=1$ super Yang-Mills theory to one dimension. 
%The index $\alpha$ of the fermionic matrices $\psi_\alpha$ corresponds to the spinor index in ten dimension, and 
%$\psi_\alpha$ is Majorana-Weyl in ten-dimensional sense. 

It is possible to turn-on a background flux, keeping 16 supersymmetries. 
(Supersymmetry transformation is modified.) 
The resulting model is called Berenstein-Maldacena-Nastase (BMN) matrix model\cite{Berenstein:2002jq}:
\begin{eqnarray}
S_{BMN}=S_{BFSS}+\Delta S, 
\label{eq:BMN_1}
\end{eqnarray}
where 
\begin{eqnarray}
\Delta S
&=&
\frac{N}{\lambda}\int dt\ {\rm Tr}\left\{
\frac{\mu^2}{2}\sum_{i=1}^3X_i^2 
+
\frac{\mu^2}{8}\sum_{a=4}^9X_a^2 
+
i\sum_{i,j,k=1}^3\mu\epsilon^{ijk}X_iX_jX_k
+
\frac{3i\mu}{4}
\bar{\psi}\gamma^{123}\psi
\right\}. 
\nonumber\\\label{eq:BMN_2}
\end{eqnarray}
Here $\epsilon_{ijk}$ is the structure constant of $SU(2)$ (i.e. $\epsilon_{123}=+1, \epsilon_{213}=-1$ etc), and hence
$i\sum_{i,j,k=1}^3\mu\epsilon^{ijk}{\rm Tr}\left(X_iX_jX_k\right)=3i\mu{\rm Tr}\left(X_1,[X_2,X_3]\right)$. 
It has a fuzzy sphere solution $X_{1,2,3}=\mu J_{1,2,3}$, $[J_i,J_j]=i\epsilon_{ijk} J_k$, $X_{4,5,\cdots,9}=0$, $A_t=0$ and $\psi=0$,
which preserves the full supersymmetry. 

BFSS matrix model has been studied extensively in the past. 
Here is a list of previous numerical studies:
\begin{itemize}
\item
Black hole mass at finite temperature has been studied in 
Refs.\citen{Anagnostopoulos:2007fw,Catterall:2008yz,Catterall:2009xn,
Kadoh:2015mka,Filev:2015hia,Hanada:2008ez,Hanada:2013rga,Berkowitz:2016jlq}.
In particular, the large-$N$, continuum values has been evaluated in a wide temperature region in Ref.\citen{Berkowitz:2016jlq}, 
and the leading supergravity prediction $7.4N^2\lambda^{-3/5} T^{14/5}$ has been confirmed by a three parameter fit with 
$a \lambda^{-3/5} T^{14/5} + b \lambda^{-6/5} T^{23/5} + c \lambda^{-8/5} T^{29/5}$. 
The stringy $\alpha'$ corrections has been studied in Ref.\citen{Hanada:2008ez} and Ref.\citen{Kadoh:2015mka}, 
with taking into account only the next-to-leading term $\lambda^{-6/5} T^{23/5}$, and different results have been obtained. 
More precise analysis has been performed in Ref.\citen{Berkowitz:2016jlq}, by taking the large-$N$ and continuum limits. 
It turned out that the next-to-next-to-leading term $\lambda^{-8/5} T^{29/5}$ is needed in the temperature region considered in 
Ref.\citen{Hanada:2008ez} and Ref.\citen{Kadoh:2015mka}, 
and with that term, gauge theory and string theory are consistent.  
The $g_s$ correction has been studied in Ref.\citen{Hanada:2013rga} and Ref.\citen{Berkowitz:2016jlq}, 
and agreed with dual gravity results\cite{Hyakutake:2013vwa,Hyakutake:2014maa} quantitatively. 
So far, these simulations provide us with the strongest evidence supporting the gauge/gravity duality at stringy level.

\item
The Wilson loop has been calculated\cite{Hanada:2008gy} and the dual gravity prediction has been reproduced. 

\item
Several two-point functions are studied\cite{Hanada:2009ne,Hanada:2011fq} and 
the generalization of GKPW relation\cite{Witten:1998qj,Gubser:1998bc} to D0-brane theory\cite{Sekino:1999av}
has been justified. 

\item
Mean-field method has also been applied to this theory\cite{Kabat:1999hp,Kabat:2000zv,Kabat:2001ve,Iizuka:2001cw}. 
This method may work better when the sign problem and/or instability become severe. 

\end{itemize}

The next targets include:
\begin{itemize}
\item
M-theory parameter region. In particular, can the Gregory-Laflamme transition\cite{Hyakutake:2015rqa} be seen? 
Can the Schwarzschild black hole in M-theory be described? 

\item
The trivial vacuum has a large-$N$ suppressed instability\cite{de Wit:1988ct}, 
which can be regarded\cite{Hanada:2013rga,Berkowitz:2016znt} as the Hawking radiation\cite{Hawking:1976ra} of D0-branes.  
By determining the potential with a probe D0-brane, it should be possible to read off the black hole geometry from matrices.
It can answer a long-standing problem -- the bulk geometry from gauge theory -- and 
should provide us with valuable insights into black hole information puzzle.    
Earlier attempts with essentially the same ideas can be found in Refs.\citen{Iizuka:2001cw,Ferrari:2012nw}.

\item
There are dual gravity predictions for BMN matrix model in trivial vacuum\cite{Costa:2014wya}. Can they be reproduced? 

\item
It has been proposed that the six-dimensional superconformal theory describing M5-brane
is equivalent to the BMN matrix model about certain fuzzy sphere background\cite{Maldacena:2002rb}. 
(See Ref.\citen{ArkaniHamed:2001ca} for another interesting proposal.)
It is possible to test this proposal, by comparing the finite-temperature 
free energy with eleven-dimensional supergravity on AdS$_7\times$S$^4$.

\end{itemize}

%%%%%%%%%%%%%%%%%
%%%%%%%%%%%%%%%%%
\section{Monte Carlo study with imaginary time}\label{sec:imaginary_time_simulation}	
%%%%%%%%%%%%%%%%%
%%%%%%%%%%%%%%%%%

%%%%%%%%%%%%%%%%%
%%%%%%%%%%%%%%%%%
\subsection{Regularization methods}	
%%%%%%%%%%%%%%%%%
%%%%%%%%%%%%%%%%%

%%%%%%%%%%%%%%%%%
%%%%%%%%%%%%%%%%%
\subsubsection{Lattice approach}\label{sec:lattice}	
%%%%%%%%%%%%%%%%%
%%%%%%%%%%%%%%%%%
Supersymmetry algebra contains the infinitesimal translation, $\{Q_\alpha, Q_\beta\} \sim \gamma_{\alpha\beta}^\mu\partial_\mu$. 
Therefore, lattice explicitly breaks supersymmetry by definition. 
Even if the tree-level action preserves supersymmetry, the radiative correction can give 
supersymmetry-breaking terms which lead to a wrong continuum limit. 

One approach to circumvent this problem is to use other symmetries to forbid supersymmetry-breaking radiative corrections. 
In the case of 4d minimal SYM (4d YM + gluino), the only possible SUSY-breaking radiative correction 
is the mass of gluino. This can also be forbidden by chiral symmetry. Therefore, by keeping chiral symmetry on lattice, 
correct supersymmetric continuum limit can be realized\cite{Kaplan:1983sk,Curci:1986sm}. 
Actual simulations based on this idea have been tried by a few groups\cite{Giedt:2008xm,Endres:2009yp,Kim:2011fw}. 
(It would also be possible to use other actions which do not respect chiral symmetry, 
and tune the gluonino mass to zero\cite{Demmouche:2010sf,Bergner:2013nwa,Bergner:2015adz}.)  
This method does not work any more, once scalar fields appear, i.e. in extended supersymmetric theory and supersymmetric QCD,  
because the only symmetry which can forbid scalar mass is supersymmetry itself! 

Although supersymmetry can never be fully preserved on lattice, it is still possible to keep sub-algebra 
which does not contain infinitesimal translation. 
If it is done in a clever manner, a part of R-symmetry can also be kept. 
If these symmetries are good enough to control the radiative corrections, the right continuum limit can be obtained\cite{Kaplan:2002wv}. 
Various lattice models based on this idea have been formulated\cite{Cohen:2003xe,Cohen:2003qw,
Kaplan:2005ta,Sugino:2003yb,Sugino:2004qd,Sugino:2004uv,Catterall:2004np,Catterall:2005fd,D'Adda:2005zk}. 
It turned out this strategy works in $(0+1)$-d and $(1+1)$-d, to all order in perturbation theory. 
In higher dimensions, the exact symmetries are not powerful enough to exclude possible fine tunings. 
2d ${\cal N}=(2,2)$ SYM, which is the dimensional reduction of 4d ${\cal N}=1$ SYM, has been studied numerically
\cite{Kanamori:2008bk,Hanada:2009hq,Hanada:2010qg} and the validity of the regularizations at nonperturbative level has been confirmed. 
2d ${\cal N}=(8,8)$ SYM, which is dual to the D1-brane theory, has also been studied\cite{Catterall:2010fx}. 

In order to study 4d theory without fine tunings, it is necessary to use matrix model methods explained below. 
Still, it might be possible to use lattice and perform parameter fine tunings. 
Recent attempts along this direction can be found in \cite{Catterall:2014vka,Catterall:2015ira,Schaich:2015daa}. 

One remark is in order, before concluding this section.
In the literature on lattice supersymmetry, a word `topological twist' is often used, 
and many people mistakenly think these lattice theories describe topological field theories. 
This is just a bad, misleading choice of a technical terminology; these theories do describe usual SYM theories.

%%%%%%%%%%%%%%%%%
%%%%%%%%%%%%%%%%%
\subsubsection{Matrix model methods (1): noncommutative space}\label{sec:NCYM}	
%%%%%%%%%%%%%%%%%
%%%%%%%%%%%%%%%%%
Lattice looks like a very natural regularization to us. 
However what looks natural to human eyes is not necessarily always the best choice. 
For theories which have natural realization in string theory, we can use a better strategy based on {\it noncommutative space}. 
The key is the Myers effect\cite{Myers:1999ps}, explained in Sec.~\ref{sec:D2/M2} -- with suitable background flux, 
D$p$-branes spread out and form 
a kind of `lattice,' which can be regarded as D$(p+2)$-brane. 
In terms of gauge theory, two-dimensional `lattice' can be embedded in scalar fields in $(p+1)$-dimensional gauge theory, 
so that $(p+3)$-dimensional theory is obtained. 
Because this `lattice' has a physical realization in string theory, it behaves much better than man-made lattices. 

As a concrete example, let us consider a system of $N$ D0-branes, which is described by 
U($N$) maximally supersymmetric matrix quantum mechanics. 
It is possible to introduce background flux without breaking supersymmetry. The action changes to BMN \cite{Berenstein:2002jq}.  
It has a fuzzy sphere solution, which keeps full 16 supersymmetry. By expanding the theory around $k$-coincident 
fuzzy sphere background, $(1+2)$-dimensional super Yang-Mills on fuzzy sphere is obtained\cite{Maldacena:2002rb}.  
I refer the detail of the embedding of the noncommutative space to matrices to original references, e.g.\cite{Aoki:1999vr,Iso:2001mg}. 
Historically, Yang-Mills theory on noncommutative space has appeared in a rather different context, 
as the Twisted Eguchi-Kawai model\cite{GonzalezArroyo:1982hz}, which is explained in Sec.~\ref{sec:Eguchi_Kawai}. 
Later it has been re-interpreted in the context of string theory. 

About $k$-coincident fuzzy sphere background (i.e. $k$ copies of spin $s$ representations, 
$(2s+1)k=N$), $(1+2)$-d U($k$) SYM and $(0+1)$-d U($N$) are related as\cite{Maldacena:2002rb} 
\begin{eqnarray}
\Lambda_{\rm UV}\sim\mu s,
\qquad
\Lambda_{\rm IR}\sim\frac{1}{R_{S^2}}\sim \mu,
\qquad
 \Theta\sim\frac{1}{\mu^2 s}, 
 \qquad
 g_{(2+1)}^2\sim \Theta g_{(0+1)}^2, 
\end{eqnarray}
where $\Theta$ is the noncommutativity parameter. 
$(1+2)$-d U($k$) SYM on commutative space is obtained by sending $\Lambda_{\rm UV}\to\infty$ and $\Theta\to 0$. 
Note that the commutative limit $\Theta\to 0$ is singular for generic quantum field theories, 
due to a mixing between ultraviolet and infrared degrees of freedom \cite{Minwalla:1999px}. 
Fortunately, the commutative limit is not singular for maximally supersymmetric theories\cite{Hanada:2014ima,Matusis:2000jf}. 
For this reason, supersymmetry is essential for this construction. Supersymmetry is needed also for the stability 
of the background matrix configurations\cite{Azeyanagi:2007su,Azeyanagi:2008bk}. 

In order to formulate the four-dimensional theory, we can start with $(1+1)$-dimensional super Yang-Mills with 
a BMN-like deformation. 
By expanding the theory about fuzzy sphere background, it is possible to obtain $(3+1)$-dimensional theory. 
Lattice regularization of such $(1+1)$-dimensional theory can be obtained\cite{Hanada:2010kt,Hanada:2010gs,Kato:2011yh} 
by modifying the theories explained in Sec.~\ref{sec:lattice}. It is possible to show that the parameter fine-tuning is not needed 
if one takes the continuum limit of the lattice direction first and then take the large-$N$ limit\cite{Hanada:2010kt,Hanada:2010gs}.

%%%%%%%%%%%%%%%%%
%%%%%%%%%%%%%%%%%
\subsubsection{Matrix model methods (2): Eguchi-Kawai reduction}\label{sec:Eguchi_Kawai}	
%%%%%%%%%%%%%%%%%
%%%%%%%%%%%%%%%%%
The Eguchi-Kawai equivalence\cite{Eguchi:1982nm} states that large-$N$ matrix models 
about certain background matrix configurations become equivalent to large-$N$ quantum field theories 
in higher dimension. For example, by taking the limit of infinitely large noncommutativity parameter $\Theta\to\infty$ 
of the fuzzy sphere background in the BMN matrix model (see Sec.~\ref{sec:NCYM}), 
only the planar sector survives\cite{GonzalezArroyo:1982hz}, and hence it is equivalent to 
the $(1+2)$-dimensional planar SYM. This is the twisted Eguchi-Kawai reduction\cite{GonzalezArroyo:1982hz}. 
There exists a formulation\cite{Ishii:2008ib} of 4d ${\cal N}=4$ SYM on $S^3$ 
based on the Eguchi-Kawai reduction; 
it uses a certain multi fuzzy sphere background of the BMN matrix model. Two dimensions are generated by fuzzy sphere 
and the other dimension is generated via the quenched Eguchi-Kawai reduction\cite{Bhanot:1982sh,Gross:1982at,Parisi:1982gp}. 
The same method can also be used to construct other theories\cite{Hanada:2009hd}, 
including the ABJM theory and supersymmetric QCD. 

In order to prove the Eguchi-Kawai equivalence, it is necessary to assume two conditions, 
(i) the stability of the background matrix configuration, and (ii) dominance of the planar sector.  
In supersymmetric theories, the former is automatic unless the supersymmetry is broken 
by the thermal boundary condition. The latter is satisfied in the 't Hooft large-$N$ limit 
($g_{YM}^2N$ fixed). Somewhat interestingly,  
there is evidence that it can also work in the M-theory limit, in which $g_{YM}^2N$ grows with $N$\cite{Azeyanagi:2012xj}.

%%%%%%%%%%%%%%%%%
%%%%%%%%%%%%%%%%%
\subsubsection{Momentm cutoff method for $(0+1)$-d theories}	
%%%%%%%%%%%%%%%%%
%%%%%%%%%%%%%%%%%
For quantum mechanics (i.e. $(0+1)$-d theory), the momentum cutoff method\cite{Hanada:2007ti} is applicable. 
Let us consider the BFSS matrix model \eqref{eq:BFSS} as a concrete example. 
Firstly, the gauge symmetry is fixed. Here we take the static diagonal gauge, 
\begin{eqnarray}
A_t=\frac{1}{\beta}\cdot{\rm diag}(\alpha_1,\cdots,\alpha_N),
\qquad
-\pi<\alpha_i\le\pi. 
\end{eqnarray} 
Associated with this gauge fixing, we add the Faddeev-Popov term 
\begin{eqnarray}
S_{F.P.}
&= &
-
\sum_{i<j}2\log\left|\sin\left(\frac{\alpha_i-\alpha_j}{2}\right)\right|
\label{eq:Faddeev-Popov}
\end{eqnarray}
to the action.
Then the fields are expanded in Fourier modes, and the momentum cutoff is introduced. 
Although the momentum cutoff breaks supersymmetry explicitly,
supersymmetry-breaking radiative corrections do not appear, due to the absence of the ultraviolet divergence. 
This method has been used in 
Refs.\citen{Anagnostopoulos:2007fw,Hanada:2008gy,Hanada:2008ez,Hanada:2013rga,Hanada:2011fq,Hanada:2009ne}. 

%%%%%%%%%%%%%%%%%
%%%%%%%%%%%%%%%%%
\subsection{Some technicalities}	
%%%%%%%%%%%%%%%%%
%%%%%%%%%%%%%%%%%

%%%%%%%%%%%%%%%%%
%%%%%%%%%%%%%%%%%
\subsubsection{Dynamical fermion, sign problem and RHMC}	
%%%%%%%%%%%%%%%%%
%%%%%%%%%%%%%%%%%
In order to study the maximally supersymmetric Yang-Mills with the dynamical fermion, 
we need to integrate out fermions like in lattice QCD. 
Then the Pfaffian, which is the square root of the determinant up to a sign, appears because 
the fermion is Majorana-Weyl. If we write the fermionic part to be $S_{\rm F}=\bar{\psi}D\psi$, 
\begin{eqnarray}
\int [dA][dX][d\psi]e^{-S_{\rm B}-S_{\rm F}}
=
\int [dA][dX]{\rm Pf}D(A,X)\ e^{-S_{\rm B}}. 
\end{eqnarray}
In general, ${\rm Pf}D(A,X)$ is complex, and hence the standard importance sampling method is not applicable. 
One way to circumvent this {\it sign problem} is to use phase-quenched ensemble with the weight 
$|{\rm Pf}D(A,X)|\cdot e^{-S_{\rm B}}$
and take into account the phase factor via the phase reweighting method, 
 \begin{eqnarray}
 \langle
 O(A,X)
 \rangle_{\rm full}
 =
 \frac{ \langle
 O(A,X)\cdot e^{i\theta}
 \rangle_{\rm phase\ quench}}{
 \langle
e^{i\theta}
 \rangle_{\rm phase\ quench}
 }. 
 \end{eqnarray}
Here $e^{i\theta}={\rm Pf}D(A,X)/|{\rm Pf}D(A,X)|$. 

The phase reweighting is very costly, because one has to calculate the pfaffian. 
In fact, however, the phase $\theta$ is very close to zero in a wide parameter region, 
and practically  
\begin{eqnarray}
 \langle
 O(A,X)
 \rangle_{\rm full}
=
 \langle
 O(A,X)
 \rangle_{\rm phase\ quench}
 \label{eq:mystery}
 \end{eqnarray}
holds within statistical error. Furthermore, as long as the phase can be calculated by brute force, 
it has been observed that \eqref{eq:mystery} always holds within error. 
The reason behind it is the absence of the correlation between the phase and observables\cite{Hanada:2011fq,Berkowitz:2016jlq}.
This has been confirmed only numerically; it should be related to supersymmetry, although the actual mechanism has not been understood yet. 
Based on this observation, previous simulations ignored the phase factor. 

The phase-quenched theory can be simulated by using the RHMC method\cite{Clark:2003na}. For details, see e.g. 
Ref.\citen{Catterall:2007fp}.

%%%%%%%%%%%%%%%%%
%%%%%%%%%%%%%%%%%
\subsubsection{Flat directions}	
%%%%%%%%%%%%%%%%%
%%%%%%%%%%%%%%%%%
Supersymmetric gauge theories have flat directions, i.e. 
eigenvalues of $X_I$ can grow indefinitely without costing potential energy, 
as long as $[X_I,X_{I'}]=0$. 
Because the distribution of the eigenvalues determines the vacuum, 
it is important to control the flat direction and stay in the correct phase during the simulation. 
Historically, it took some time until this problem has been correctly appreciated. 
Many lattice simulations at early days were trapped in unphysical phases, 
and many people suspected that 
the lattice formulations, including the ones which had been proven to work to all order in perturbation,
could fail at nonperturbative level. 
This problem has been pointed out in Ref.\citen{Hanada:2010qg} and the standard recipe to solve 
this problem has been provided there. 

Typically, the flat direction disappears at large $N$ and/or large volume. 
When one is interested in the $1/N$-correction, somehow one has to cope with the flat direction, 
because the existence of the flat direction itself is an important feature of the correction. 
A recipe for this case can be found in Ref.\citen{Hanada:2013rga}.

%%%%%%%%%%%%%%%%%
%%%%%%%%%%%%%%%%%
\section{Real-time study}\label{sec:real_time}
%%%%%%%%%%%%%%%%%
%%%%%%%%%%%%%%%%%
The real-time dynamics is even more interesting than the imaginary time properties. 
For example, real-time dynamics of the Hawking radiation should be an important piece of the complete solution of the information puzzle. 
Also, if we can see the moment of the topology change from a black string to a black hole, it should be possible to learn
intuition into quantum gravitational effects. 
However, the real-time simulation of a quantum theory is notoriously difficult. How can we possibly do it?

Here I present some arguments which suggest that {\it quantum gravity can be much easier than QCD, when it comes to real-time dynamics}. 
The reason is that `classical' and `quantum' in the gravity and gauge theory sides are different notions. 
It can be understood by recalling the reason that many people use the gauge/gravity duality to learn about quantum field theory; 
{\it classical gravity corresponds to quantum field theory}. Just in the same manner, {\it classical field theory already 
knows quantum gravity} to some extent. 
There are two kinds of stringy effects, $\alpha'$ and $g_s$. 
Weak coupling in gauge theory side means large $\alpha'$ correction. 
We can still tune $1/N$, which corresponds to $g_s$. 
Below I will argue that the classical field theory is already useful for learning about the thermalization of a black hole 
and black hole/black string transition. By taking into account a small fraction of quantum effects in the gauge theory side, 
evaporation of a black hole can be studied. Then I will comment on quantum simulator.  

%%%%%%%%%%%%%%%%%
%%%%%%%%%%%%%%%%%
\subsection{Classical simulation}	
%%%%%%%%%%%%%%%%%
%%%%%%%%%%%%%%%%%

%%%%%%%%%%%%%%%%%
%%%%%%%%%%%%%%%%%
%\subsubsection{D0-brane matrix mechanics and black hole}	
%%%%%%%%%%%%%%%%%
%%%%%%%%%%%%%%%%%

Let us start with the D0-brane matrix quantum mechanics. At high temperature, 
it reduces to the classical matrix mechanics. The fermion drops off, 
and the Lagrangian becomes 
\begin{eqnarray}
L=
\frac{1}{g_{YM}^2}{\rm Tr}
\left(
\frac{1}{2}(D_t X_M)^2 
+
\frac{1}{4}[X_M,X_N]^2\right). 
\end{eqnarray}
This model exhibits classical chaos\cite{Savvidy:1982jk}. Regardless of the detail of the initial condition, 
generic initial conditions evolve to the `typical state,' which is a black hole. 

A black hole is described by a bunch of D0-branes and open strings. Suppose diagonal elements (D0-branes) are far separated. 
Then the off-diagonal elements (open strings) cannot be excited, because they cost energy from ${\rm tr}[X_M,X_N]^2$ term  
(i.e. open strings become longer and hence heavier).
When the D0-branes come closer, open strings become lighter and can be excited, and hence the dynamical degrees of freedom increase. 
Hence a black hole is entropically favored. This mechanism works both at classical and quantum levels of the matrix model. 
In the classical limit, detailed numerical experiment is possible\cite{Asplund:2011qj,Asplund:2012tg,Aoki:2015uha}. 
  
One of the important characterizations of a black hole is that it scrambles information extremely fast; 
indeed, it is conjectured to be the fastest information scrambler \cite{Sekino:2008he}. 
Then, because of the gauge/gravity duality, the matrix quantum mechanics must also be the fastest scrambler. 
In terms of the matrix model, the Lyapunov exponent $\lambda_L$, which is defined by the exponential growth of the perturbation 
($|\delta X|\propto e^{\lambda_L t}$), should be of order $N^0$ and a small perturbation should spread to the entire system 
in the scrambling time $t_s\sim \frac{\log N}{\lambda_L}$.  It has further been argued that the Lyapunov exponent must satisfy 
$\lambda_L\le 2\pi T$ and the bound is saturated at the strong coupling limit \cite{Maldacena:2015waa}. 

The Lyapunov behavior of the matrix model has been studied in the classical limit.\cite{Aref'eva:1998mk,Gur-Ari:2015rcq}
(See also Ref.\citen{Tsukiji:2015rra} for an attempt to include quantum effects.)
It has been found\cite{Gur-Ari:2015rcq} that the Lyapunov exponent is of order $N^0$ and satisfies the proposed bound, 
and the entire Lyapunov spectrum has a nice large-$N$ limit which is consistent with the argument in the gravity side. 

Classical Yang-Mills theory with nonzero spatial dimension does not have a well-defined 
continuum limit, because of the ultraviolet catastrophe; there are too many ultraviolet degrees of freedom, and 
energy and momentum continue to flow to ultraviolet indefinitely. 
(Note that this is exactly the reason why Planck had to introduce $\hbar$.)
Still, if the initial condition is chosen so that only infrared modes are excited, 
we can expect that an approximate thermalization takes places and then the ultraviolet catastrophe turns on 
only at a later stage (see \citen{Kurkela:2012hp} for a study along this direction). 
Then, the classical simulation can be justified as long as we do not see very late time. 
Among various possible directions, the chaos and black hole/black string transition would be of particular interest. 
As for the former, several groups have already studied SU($2$) and SU($3$) theories 
(see e.g. Refs.\citen{Kunihiro:2008gv,Kunihiro:2010tg}). 
It would be interesting to study larger $N$ and compare the results with the dual gravity description of $4d$ ${\cal N}=4$ SYM. 
As for the latter, it should be possible to see the moment of the topology change, which is not accessible with any other methods. 
It should tell us valuable hints, e.g. the actual final state and possible intermediate states, (un-)importance of the $1/N$-correction, 
the time scale for the topology change. 
A recipe to extract the topology from matrices is given in Ref.\citen{Azeyanagi:2009zf}; see also refs.\citen{Schneiderbauer:2016wub,Ishiki:2015saa}. 

%%%%%%%%%%%%%%%%%
%%%%%%%%%%%%%%%%%
\subsection{Quantum effects}	
%%%%%%%%%%%%%%%%%
%%%%%%%%%%%%%%%%%
By taking the quantum effects into account, 
the evaporation of a black hole (black zero-brane) can be described as follows \cite{Berkowitz:2016znt}.  
Even at high temperature, when a D0-brane goes sufficiently far from the black hole, quantum effect becomes non-negligible
and then the flat direction opens up. Once a D0-brane reaches this distance, it can travel almost freely away from the BH. 
It is the Hawking radiation of a D0-brane. A simple entropy argument shows that 
the black hole can evaporate completely, by emitting D0-branes one-by-one. 
During this evaporation process, temperature of the black hole goes up.
Hence, an important prediction by Hawking -- negative specific heat of a black hole -- naturally follows from 
the gauge theory description. 
In fact, the negative specific heat can be shown analytically, even at low temperature where 
the classical description is not valid at all\cite{Berkowitz:2016znt}. 
Therefore, the classical simulations explained in the previous section are justified for sufficiently old black holes.

So far, all these arguments\cite{Berkowitz:2016znt} are done analytically. 
With numerical methods, we can see the quantitative detail of the black hole evaporation. 
For example, by the imaginary time simulation, it is possible to determine the potential which determines 
the motion of emitted D0-branes. 
In gravity language, it translates to the metric of the black hole solution. 
By tuning the coupling constant and $N$, we can see how the stringy effects modify the metric. 
Furthermore, by using the metric obtained in this manner, the black hole evaporation can be simulated numerically. 

The quantum effects would be important also in the study of Yang-Mills theory with nonzero spatial dimension, 
especially because it would allow us to dial the $\alpha'$-correction. Although it is challenging, 
it should be easier than similar calculations for QCD, for the reason explained in the beginning of this section. 

%%%%%%%%%%%%%%%%%
%%%%%%%%%%%%%%%%%
\subsection{Quantum simulator}	
%%%%%%%%%%%%%%%%%
%%%%%%%%%%%%%%%%%
Recently, the quantum simulation based on Atomic/Molecular/Optical (AMO) physics are pursued by many physicists. 
The idea is to realize theoretically interesting systems, such as the Hubbard model, experimentally, 
for example by trapping atoms on an optical lattice, and do an actual, rather than numerical, experiment. 
Several groups are trying to design quantum simulator for lattice QCD; see \cite{Wiese:2013uua} for a review. 

Seen from string theorists, this approach is of particular interest not just for theoretical reason; 
{\it it would allow us to make a black hole in a laboratory}.    
If one can realize a non-gravitational theory which is dual to a black hole, 
it is equivalent to creating an actual black hole! 
Then it should be possible to see how a small, 
quantum black hole forms and evaporates; it would be possible to study quantum gravity experimentally. 

At this moment, it is not easy to construct such field theories, e.g. SYM, experimentally. 
However a simpler system, for example the Sachdev-Ye-Kitaev model\cite{Sachdev:2015efa}, 
would be within reach\cite{danshita}. At very least, such approach would be much more tractable than 
collider experiment at Planck scale. It would be nice if lattice theorists and AMO physicists could make 
string theory and quantum gravity {\it experimental} science. 

Note that such theories may or may not be {\it the} theory which describes our world.   
Still, by studying various quantum gravitational systems, it would be possible to understand universal and non-universal features.  
Perhaps we find a universal problem which falsifies the string theory. 
Perhaps we arrive at a better understanding about {\it the} quantum gravity describing our world. 
Or -- although it would be a very speculative statement -- if we are actually living in the string theory multiverse, then it might be possible to create 
{\it other universes} in {\it our universe}.

%%%%%%%%%%%%%%%%%
%%%%%%%%%%%%%%%%%
\section{Conclusion and discussions}	
%%%%%%%%%%%%%%%%%
%%%%%%%%%%%%%%%%%
In this review, I tried to convince lattice and string theorists that lattice method is an important piece of the study of superstring/M-theory, 
especially on its quantum gravitational aspects. 
The reason is simple: superstring/M-theory is the most promising candidate of 
the theory of quantum gravity, and lattice gauge theory is the only tool available at this moment 
to solve many of the important problems there. 
The developments in these ten years enabled us to study most supersymmetric gauge theories 
relevant for superstring/M-theory numerically. 
By now, except for sociological reasons (i.e. lack of mutual understanding and funding), 
nothing prevents collaboration between lattice and string theorists. 
And such sociological factors can easily be resolved as soon as lattice and string theorists start talking to each other. 

There are many interesting topics which are not covered by this review, for example:
\begin{itemize} 
\item
The loop formulation of SYM and simulation at each fixed fermion number\cite{Steinhauer:2014oda,Bergner:2015ywa}.  
It would be useful especially when the sign problem becomes severe. 

\item
Recent developments in conformal bootstrap\cite{Rattazzi:2008pe,ElShowk:2012ht}. 

\item
Hamiltonian truncation method.\cite{Hogervorst:2014rta} 
A related method applied to matrix quantum mechanics can be found in Refs.\citen{Ambrozinski:2014oka,Campostrini:2004bs}. 

\item
Holographic cosmology\cite{McFadden:2009fg}, which can make nontrivial prediction on cosmological observables combined with lattice simulation.

\item
A matrix model description of type IIB superstring.\cite{Ishibashi:1996xs}

\item
A lattice study of the Green-Schwarz action of string worldsheet.\cite{Forini:2016sot}

\end{itemize}

Once many lattice theorists start working in interesting directions related to superstring/M-theory, 
it should motivate string theorists to make 
even more interesting proposals, and the research field of lattice gauge theory will be substantially expanded. 
Also, if string theorists realize the power of lattice and work with lattice theorists, 
they can attack challenging problems which they never imagined to be within their reaches. 
I hope that this review serves as a modest step toward fruitful collaborations between lattice and string theorists. 

\begin{center}
{\bf Acknowledgments}
\end{center}
The work of M.~H. is supported in part by the Grant-in-Aid of the Japanese Ministry 
of Education, Sciences and Technology, Sports and Culture (MEXT) 
for Scientific Research (No. 25287046). 
%%%%%%%%%%%%%%%%%
%%%%%%%%%%%%%%%%%
%%%%%  Reference %%%%%%
%%%%%%%%%%%%%%%%%
%%%%%%%%%%%%%%%%%
%\begin{thebibliography}{000} %for 3 digits
%\begin{thebibliography}{00}  %for 2 digits

\end{document}